\documentclass[a4paper,11pt]{article}
\pdfoutput=1
\usepackage{jheppub}
\usepackage{mathrsfs}

\let\originalleft\left
\let\originalright\right
\renewcommand{\left}{\mathopen{}\mathclose\bgroup\originalleft}
\renewcommand{\right}{\aftergroup\egroup\originalright}

\newcommand{\deriv}[3][]{\frac{\d^{#1}#2}{\d#3^{#1}}}

\newcommand{\pd}{\partial}
\renewcommand{\t}[1]{\text{#1}}
\renewcommand{\d}{\t{d}}
\newcommand{\e}{\t{e}}
\renewcommand{\i}{\t{i}}
\newcommand{\abs}[1]{\left|#1\right|}
\newcommand{\non}{\nonumber}

\newcommand{\ket}[1]{\left|#1\right\rangle}

\newcommand{\Braket}[3]{\left\langle#1\middle|#2\middle|#3\right\rangle}
\newcommand{\ev}[1]{\left\langle#1\right\rangle}
\newcommand{\lp}{\left(}
\newcommand{\rp}{\right)}
\newcommand{\lb}{\left[}
\newcommand{\rb}{\right]}
\newcommand{\lc}{\left\{}
\newcommand{\rc}{\right\}}
\def\bal#1\eal{\begin{align}#1\end{align}}
\newcommand{\beq}{\begin{equation} }
\newcommand{\eeq}{\end{equation} }
\newcommand{\bc}{\begin{center} }
\newcommand{\ec}{\end{center} }
\newcommand{\bfig}{\begin{figure}}
\newcommand{\efig}{\end{figure}}
\def\bmult#1\emult{\begin{multline}#1\end{multline}}
\DeclareMathOperator{\re}{Re}

\DeclareMathOperator{\sech}{sech}
\DeclareMathOperator{\csch}{csch}

\DeclareMathOperator{\am}{am}
\DeclareMathOperator{\sn}{sn}
\DeclareMathOperator{\cn}{cn}
\DeclareMathOperator{\dn}{dn}
\DeclareMathOperator{\pq}{pq}

\DeclareMathOperator{\scc}{sc}

\newcommand{\tx}{\tilde{x}}

\title{Non-Thermal Behavior in Conformal Boundary States}
\author{Kevin Kuns and Donald Marolf}
\affiliation{Department of Physics, University of California,\\
Santa Barbara, CA 93106, U.S.A.}

\emailAdd{kuns@physics.ucsb.edu}
\emailAdd{marolf@physics.ucsb.edu}

\abstract{Cardy has recently observed that certain carefully tuned states of 1+1 CFTs on a timelike strip are periodic with period set by the light-crossing time. The states in question are defined by Euclidean time evolution of conformal boundary states associated with the particular boundary conditions imposed on the edges of the strip.  We explain this behavior, and the associated lack of thermalization,  by showing that such states are Lorentz-signature conformal transformations of the strip ground state. Taking the long-strip limit implies that states used to model thermalization on the Minkowski plane admit non-thermal conformal extensions beyond future infinity of the Minkowski plane, and thus retain some notion of non-thermal behavior at late times. We also comment on the holographic description of these states.}

\begin{document}

\maketitle

\section{Introduction}

The rapid change in a quantum system from a Hamiltonian $H_0$ to a Hamiltonian $H$ is known as a quantum quench.  Quantum quenches are of experimental interest since they can be studied in laboratory systems involving ultracold atoms. They are also of theoretical interest as examples of out of equilibrium quantum systems whose thermalization, or lack thereof, can be used as a first step towards understanding thermalization in more general quantum systems. Of course, a quantum system that starts in a pure state and evolves unitarily will remain in a pure state for all time. The question of whether a quantum system thermalizes can be defined as the question of whether properties of the system, such as correlation functions or entanglement entropies, eventually reach a constant ``thermal'' value with small oscillations around that value.

The problem of a quantum quench to a Hamiltonian $H$ describing a conformal field theory (CFT) has been studied in \cite{Calabrese2005, Calabrese2006, Calabrese2007, Calabrese2007b,Calabrese:2009ez,Cardy2014, Hartman2013b}, where particularly strong results are found for $1+1$ dimensional theories. In these studies, the quantum quench is modelled as an initial condition $\ket{\psi_0}$ for the state of the CFT at the time of the quench, and this state is then evolved in time with the CFT Hamiltonian.  In \cite{Calabrese2005, Calabrese2006, Calabrese2007, Calabrese2007b,Calabrese:2009ez,Cardy2014, Hartman2013b} the initial state was taken to be the Euclidean time evolution of a conformal boundary state $\ket{B}$, or equivalently that given by performing the Euclidean path integral over a strip with some conformally-invariant boundary condition $B$ on the lower edge.  The problem thus reduces to studying a CFT with a boundary, known as a boundary CFT (BCFT) \cite{Cardy2004}. For CFTs on the infinite Minkowski plane, many properties of correlation functions and entanglement entropies reproduce those of a thermal state sufficiently long after the quench \cite{Calabrese2006, Calabrese2007, Calabrese2007b,Calabrese:2009ez, Hartman2013b}.

CFTs defined either on circles or finite intervals will generally exhibit quasi-periodic behavior at large times dictated by the density of states.  But at intermediate timescales --- long compared to the light-crossing time but short compared to that above --- one may expect them to approximately thermalize as well.  It is therefore notable that on a finite strip of length $L$ with conformally invariant (and thus energy-conserving) boundary conditions $B$, \cite{Cardy2014} recently showed that certain states are exactly periodic in time with period $L$ (in units with the speed of light set to one).\footnote{\label{noBH} As a result, even though the energy of such states can be large, in a holographic context (see appendix~\ref{sec:holographic}) the bulk duals to these states never form black holes.}  These states are defined by the Euclidean path integral over a rectangle of height $h$ and length $L$ with boundary condition $B$ imposed on all sides. Conformal invariance implies that the physics is determined by the ratio $L/h$.

This exact periodicity is explained in \cite{Cardy2014} by the constraint that the initial conditions involve only descendants of the identity. This suggests, and we show below, that these states are Lorentz-signature conformal transformations of the ground state of the strip.\footnote{Exact periodicity for a related finely-tuned local quench was found in \cite{local}, where the initial conditions involve only descendants of the identity.  Our work should generalize to this case, showing that the state after the quench is again a Lorentz-signature conformal transformation of the ground state on the strip.} Below, we refer to the above states as tuned rectangle states and denote them by $\ket{L/h}_{\t{strip}}$.

This observation is our main technical result and is shown in detail in section \ref{sec:construction}.  However, we also note that the $L \rightarrow \infty$ limit of tuned rectangles gives the states on the plane that were used to study thermalization in \cite{Calabrese2005, Calabrese2006, Calabrese2007, Calabrese2007b,Calabrese:2009ez, Hartman2013b} (and which we call $|h\rangle_{\t{Mink}}$ below).  As described in section \ref{sec:minkowski}, it follows that these states are also conformally related to the ground state on the strip of width $L$ with boundary condition $B$.  In particular, this ground states provides a smooth conformal extension of $|h\rangle_{\t{Mink}}$ beyond future infinity of the Minkowski plane, and indicates that these states retain some notion of non-thermal behavior at late times. In contrast, we note that more general states admit no such smooth conformal extensions, though we leave open the question of whether some well-defined (but non-smooth) and similarly non-thermal extension might exist. We close with some final discussion in section \ref{sec:discussion}, which suggests that the existence of such conformal extensions may be related to other possible non-thermal late-time behavior of the above Minkowski states. Appendix \ref{sec:holographic} briefly comments on associated asymptotically AdS${}_3$ bulk solutions for CFTs with holographic duals and provides technical input for the argument of section \ref{sec:minkowski}.

\section{Periodicity and Tuned Rectangle States}
\label{sec:cft-states}

We now address the tuned rectangle states $\ket{L/h}_{\t{strip}}$. Section~\ref{sec:construction} describes two methods of constructing these states.  The first conformally maps the Euclidean half plane to a rectangle and Wick rotates the result.  The second makes a Lorentz-signature conformal transformation directly from the ground state on the strip. Finally, section~\ref{sec:one-point} makes the periodic nature of these states explicit by computing the one-point functions of both primary operators and the stress tensor.

\subsection{Tuned Rectangle States from Conformal Transformations}
\label{sec:construction}

\begin{figure}
\bc
\includegraphics[scale=.9]{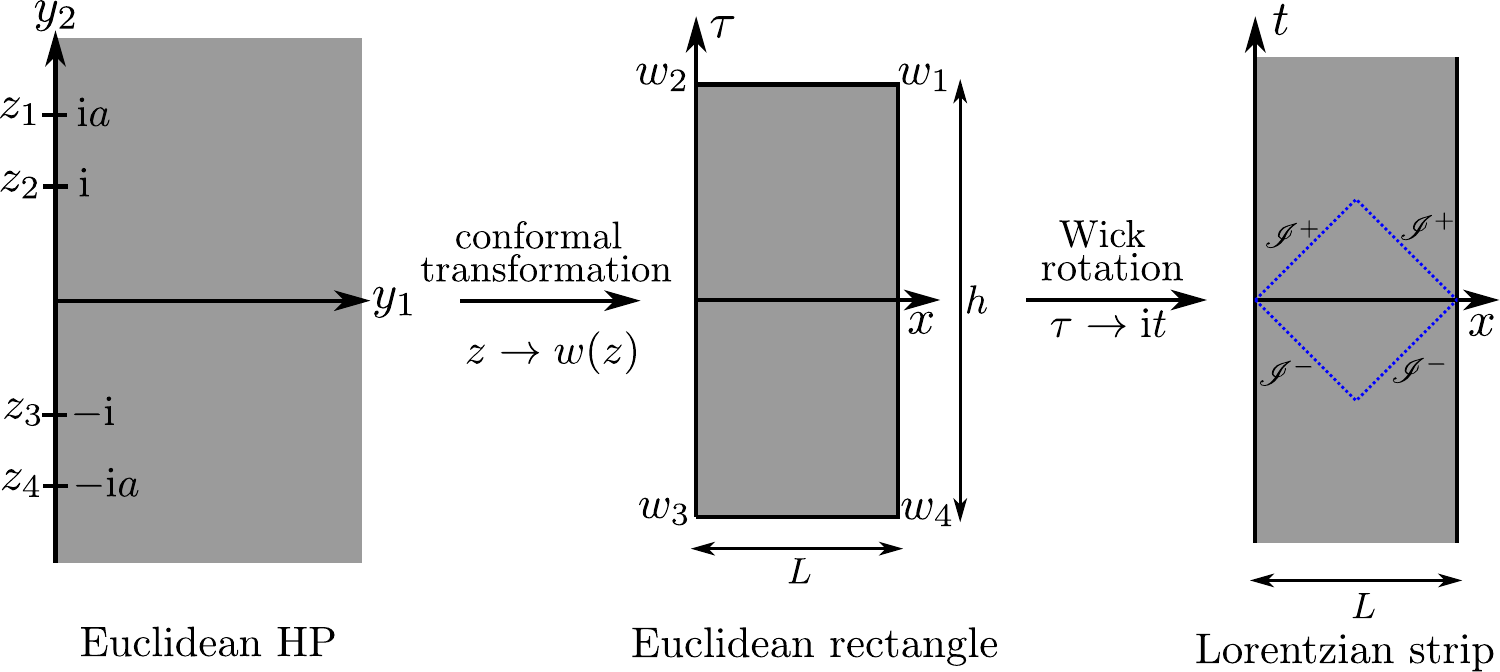}
\caption{The Lorentz-signature rectangle states can be obtained by mapping the Euclidean half-plane to a Euclidean rectangle via the conformal transformation \eqref{map} followed by a Wick rotation $\tau\to \i t$ to obtain a state defined on a Lorentzian strip. Here each $w_i$ is the image of the associated $z_i$. States defined on strips of infinite length can be obtained from a Weyl transformation that maps the dashed blue diamond, called the fundamental diamond in the text, to the Minkowski plane.}
\label{fig:map}
\ec
\end{figure}

We now show that $|L/h\rangle_{\t{strip}}$ can be built by applying conformal transformations as described above.  Our first method uses the Euclidean conformal transformation from the Euclidean half-plane.  This construction relates all Euclidean $n$-point functions to those on the half-plane, whence the Lorentz-signature correlators follow by Wick rotation.  Alternatively, we may apply the inverse transformation to relate the Euclidean path integral that computes the $t=0$ Schr\"odinger-picture wavefunctional of $|L/h\rangle_{\t{strip}}$ to a corresponding path integral on the half-plane.

The essential point is to recall that the half-plane can be conformally mapped to a rectangle using a Schwarz-Christoffel map \cite{Brown-Churchill2009}.  We choose to work with the right-half plane as shown in figure \ref{fig:map}, using Euclidean coordinates $z = y_1+\i y_2$ on the half-plane and $w = x + \i\tau$ on the rectangle. Euclidean time-reversal thus acts as $z \rightarrow \bar z, w \rightarrow \bar w$.  In particular we introduce the real parameter $a >1$ and take the points $z_1 = \i a, z_2 = \i, z_3=-\i,$ and $z_4=-\i a$ to map to the corners $w_1,w_2,w_3,$ and $w_4$ of the rectangle as shown in the figure.  So long as
\beq
\deriv{w}{z} \propto (z+\i a)^{-1/2}(z+\i)^{-1/2}(z-\i)^{-1/2}(z-\i a)^{-1/2},
\label{sc-map}
\eeq
the image of the imaginary axis will have piecewise-constant slope, changing only at the images $w_i$ of the points $z_i$ where the slope changes by $\pi/2$. The imaginary axis is thus mapped to the boundary of a rectangle. That the right-half plane is mapped to the interior of the rectangle follows by considering the image of a vector pointing from the imaginary axis between two $z_i$ into the right-half plane.

It will be useful to define $\kappa^2=1 - a^{-2}$, with $0\leq \kappa \leq 1$, so that integrating \eqref{sc-map} yields \cite{DiFrancesco-Mathieu-Senechal1997}
\beq
w(z) = - A\int_0^z \frac{\d\xi}{(\i+\xi\sqrt{1 - \kappa^2})^{1/2}(\i-\xi\sqrt{1 - \kappa^2})^{1/2}(\i+\xi)^{1/2}(\i-\xi)^{1/2}} = A F\lp \arctan z|\kappa^2\rp,
\eeq
where $F(\cdot|\cdot)$ is the incomplete elliptic integral of the first kind \cite{Abramowitz-Stegun1972} and we have inserted an arbitrary real constant $A$ to set a scale. Our parameter $\kappa$ is called $k'$ in \cite{Abramowitz-Stegun1972}.
The length and height of the rectangle are given by
\beq
L = A K\lp\kappa^2\rp \qquad \t{ and }\qquad h =2A K\lp 1- \kappa^2\rp,
\label{rectangle-dimensions}
\eeq
where $K(k)$ is the complete elliptic integral of the first kind \cite{Abramowitz-Stegun1972}. A conformal transformation to a rectangle of length $L$ is therefore accomplished by choosing $A=L/K(\kappa^2)$, or
\beq
w = \frac{L}{K(\kappa^2)} F\lp \arctan z\middle| \kappa^2\rp.
\label{map}
\eeq

\begin{figure}
\bc
\includegraphics[scale=.4]{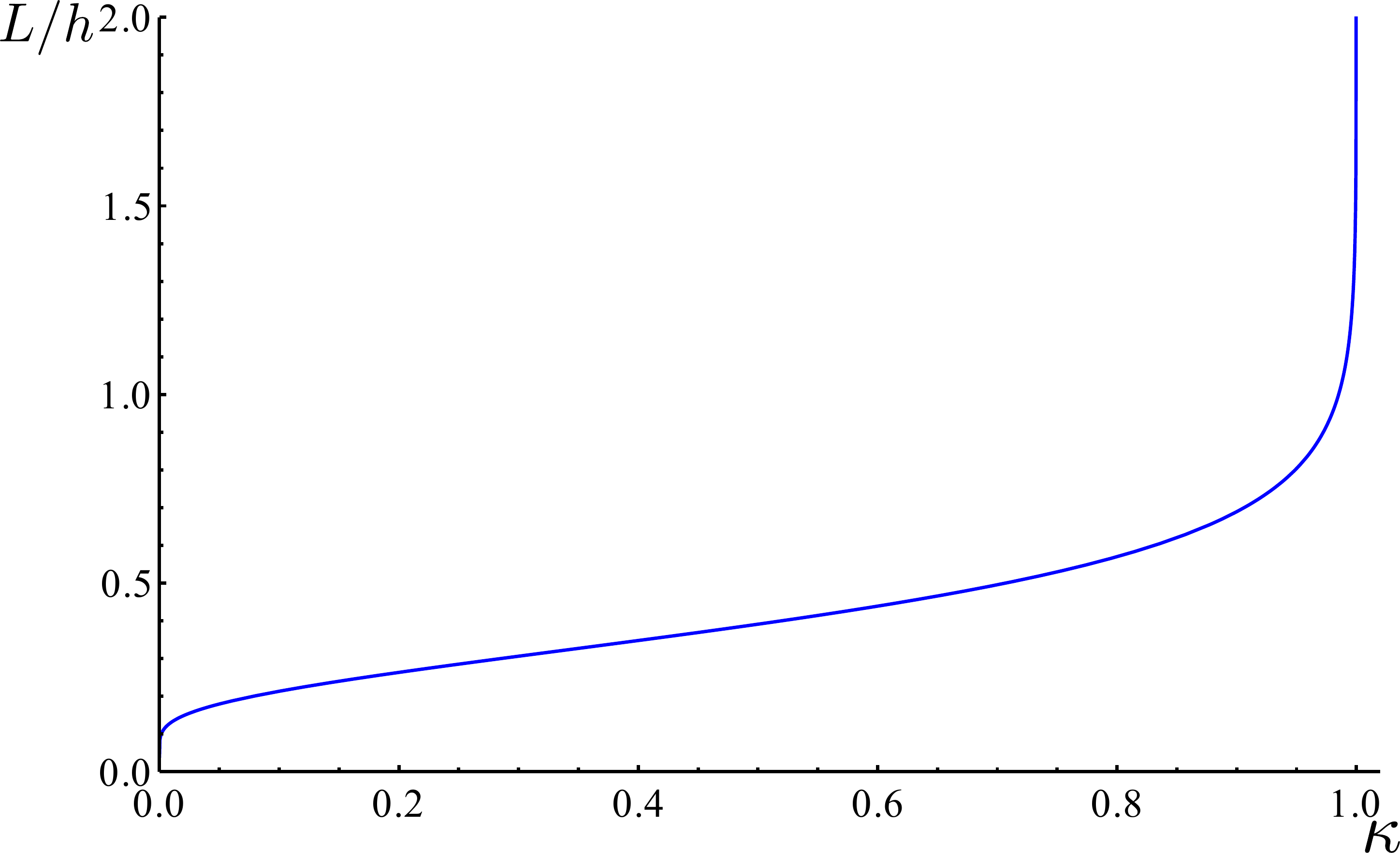}
\caption{Aspect ratio of the rectangle as a function of $\kappa$.}
\label{fig:aspect-ratio}
\ec
\end{figure}

We wish to apply this conformal transformation to a path integral on the half-plane defined by a conformally-invariant boundary condition $B$ (associated with some conformal boundary state $|B\rangle$) on the imaginary axis.  The map \eqref{map} will then relate this to a path integral on the rectangle which imposes the identical boundary condition on all four sides.  Wick rotating correlators or computing initial data at $t=0$ as described above then yields the tuned rectangle state $|L/h\rangle_{\t{strip}}$ on the Lorentzian strip with the boundary condition $B$.  For a given $B$, the set of such states is parametrized by $\kappa$, or equivalently, by $L/h$; see figure \ref{fig:aspect-ratio}. Note in particular that taking $h\to\infty$ ($\kappa \to 0$) integrates over an infinitely tall Euclidean-signature strip so that $|0 \rangle_{\t{strip}}$ is the strip ground state for boundary condition $B$.  In contrast, \cite{Cardy2014} focussed on the opposite limit $\kappa \to 1$ describing initial states with small correlation lengths.

The above construction shows that any two tuned rectangle states $\ket{L/h}_{\t{strip}}$ with the same $B$ are conformally related. In particular, taking one to be $\ket{0}_{\t{strip}}$ shows that the other (arbitrary) tuned rectangle state is a Euclidean-signature conformal transformation of the strip ground state.  Our main observation is that the time-reflection symmetry of $|L/h\rangle_{\t{strip}}$ allows us to Wick rotate this result to Lorentz-signature; i.e., any $\ket{L/h}_{\t{strip}}$ can be obtained by applying a Lorentz-signature conformal transformation to the ground state.

To see this explicitly, note that the inverse of \eqref{map} is
\beq
z = \scc \lb \frac{w}{L}K\lp \kappa^2\rp \middle|\kappa^2\rb,
\label{inverse}
\eeq
where $\scc(\cdot|\cdot)$ is a Jacobi elliptic function \cite{Whittaker-Watson1927}. Since $\scc(x|0) = \tan x$ and $K(0) = \pi/2$, setting $\kappa=0$ and taking the complex coordinate on the infinite-height Euclidean rectangle to be $w_0$ yields
\beq
z = \tan \frac{\pi w_0}{2L}.
\label{grnd-map}
\eeq
We may then map this to a Euclidean rectangle with arbitrary $\kappa$ through
\beq
w = \frac{L}{K\lp \kappa^2\rp} F\lp \frac{\pi w_0}{2L} \middle |\kappa^2\rp.
\label{conformal-euclidean}
\eeq
Analytically continuing \eqref{conformal-euclidean} implies that the general rectangle state with boundary condition $B$ can be obtained from the associated strip ground state by applying that Lorentz-signature conformal transformation
\beq
x^\pm = \frac{L}{K\lp \kappa^2\rp} F\lp\frac{\pi x_0^\pm}{2L}\middle| \kappa^2\rp,
\label{Lorentzian-map}
\eeq
in terms of null coordinates $x^\pm = x\pm t$ and $x_0^\pm$. As usual, the Wick rotation of the complex Euclidean coordinate yields two independent coordinates in Lorentz-signature.  Rather than analytically continuing the function \eqref{conformal-euclidean} in this way, some readers may prefer to think of \eqref{conformal-euclidean} as defining a reparametrization of the interval $[0,L]$ at $\tau=0$, and thus at $t=0$ of Lorentzian time.  The time-reflection symmetry then implies that the comformal map \eqref{conformal-euclidean} is fully specified by this reparametrization, and the similarly-specified Lorentz-signature conformal transformation is \eqref{Lorentzian-map}.  Thus the excited state $|L/h\rangle_{\t{strip}}$ may be written
\begin{equation}
\label{unitary}
|L/h \rangle_{\t{strip}} = U(L/h) |0 \rangle_{\t{strip}}
\end{equation}
where $U(L/h)$ is the unitary transformation that implements the symmetry of our CFT defined by \eqref{Lorentzian-map}.

Note that both $x^+-x^+_0$ and $x^--x^-_0$ are periodic with period $2L$, as is required to preserve the conformal geometry of a given strip. Since the ground state is invariant under time-translations, this observation together with the symmetry under $x \rightarrow -x$ immediately implies the result of \cite{Cardy2014} that general tuned rectangle states are also periodic with period $L$.

\subsection{One-Point Functions}
\label{sec:one-point}

The relations \eqref{unitary} and \eqref{Lorentzian-map} allow many properties of $|L/h\rangle_{\t{strip}}$ to be computed explicitly in terms of elliptic functions.  We describe expectation values of both primary operators and the stress tensor below.  In particular, the primary calculation verifies that \eqref{Lorentzian-map} reproduces the one-point functions of \cite{Cardy2014} in the relevant domain of validity.  One may also perform the computation by using \eqref{map} and Wick rotating.  For conceptual clarity we use the former method for primaries, though the anomalous term renders the latter more convenient for the stress tensor.

Consider then some primary operator $\Phi$ with scaling dimension $\Delta$.  To compute $\ev{\Phi}_{L/h} \equiv {}_{\t{strip}}\Braket{L/h}{\Phi}{L/h}_{\t{strip}}$ from \eqref{Lorentzian-map} we must first find the one-point functions in the ground state $\ev{\Phi}_0$.  For this initial step we use \eqref{grnd-map} and the one-point function on the half-plane where conformal invariance requires \cite{Calabrese2007, DiFrancesco-Mathieu-Senechal1997}
\beq
\ev{\Phi(z)}_{\t{HP}} = \frac{C_\Phi}{\lp\re z\rp^\Delta}.
\label{1p-uhp}
\eeq
Here $C_\Phi$ is a constant that depends on  the operator $\Phi$ and the boundary conditions $B$. Applying \eqref{grnd-map} yields
\beq
\ev{\Phi(w)}_0  = \abs{z'(w)}^\Delta \ev{\Phi(z)}_{\t{HP}} = C_\Phi\lb \frac{\pi}{2L}\frac{\abs{\sec^2\frac{\pi w}{2L}}}{\re\lp\tan \frac{\pi w}{2L}\rp}\rb^\Delta = C_\Phi\lp \frac{\pi}{L}\csc \frac{\pi x}{L}\rp^\Delta.
\label{1p-rect}
\eeq
As expected, the result is time-independent and thus trivial to Wick rotate to Lorentz signature.

\begin{figure}
\bc
\includegraphics[scale=.4]{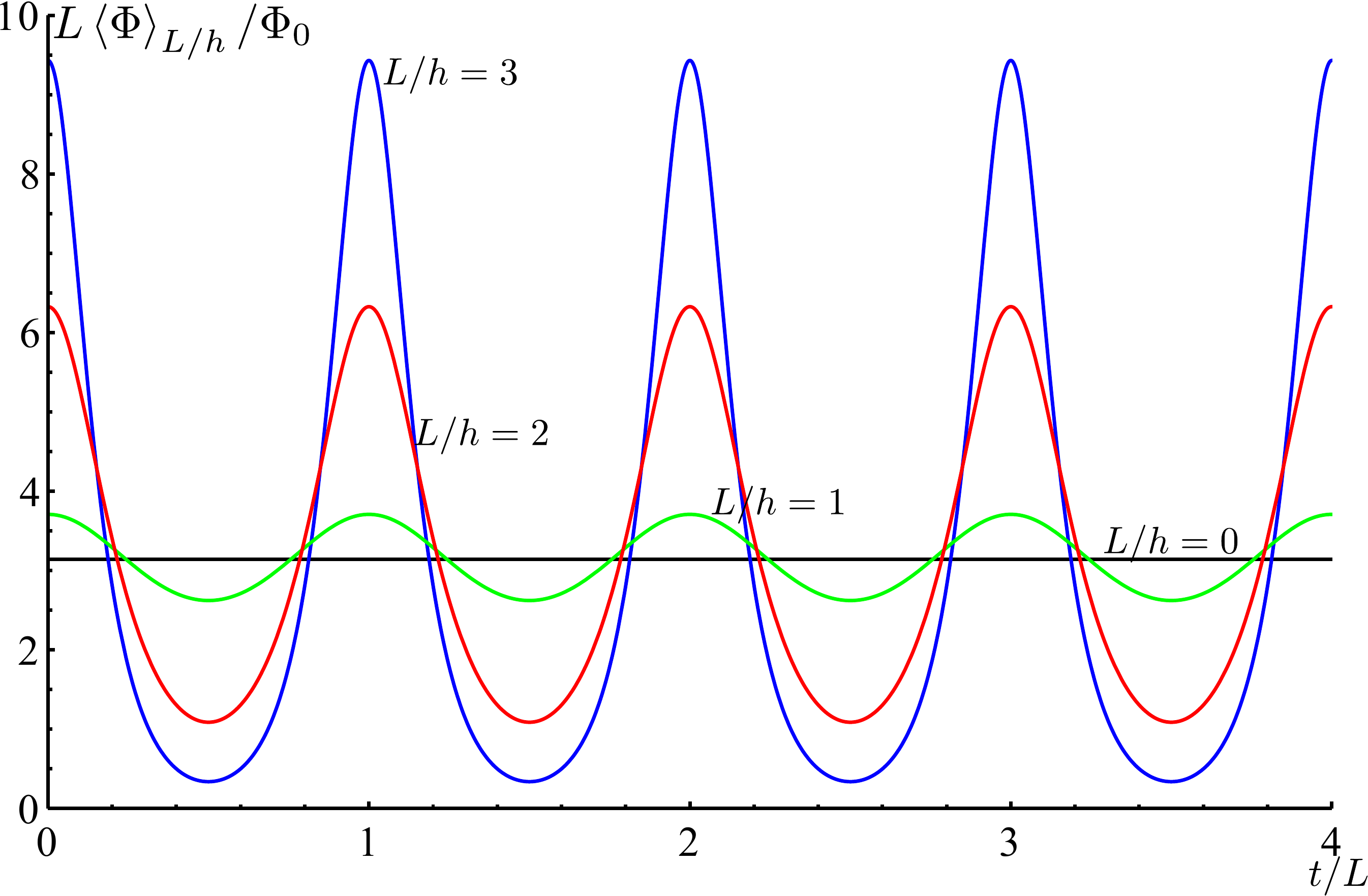}
\caption{The one-point function of a primary operator of dimension $\Delta = 1$ for different rectangle states evaluated at the center ($x=L/2$) of the Lorentzian strip.}
\label{fig:one-point}
\ec
\end{figure}

The one-point function in the general state $|L/h\rangle_{\t{strip}}$ is then obtained by applying  \eqref{Lorentzian-map}.  It is useful to note that the inverse of \eqref{Lorentzian-map} is
\beq
x_0^\pm = \frac{2L}{\pi}\am\lp \tilde{x}^{\pm}\middle| \kappa^2\rp,
\label{inverse-Lorentzian}
\eeq
where $\tilde{x}^\pm \equiv x^\pm K\lp \kappa^2\rp/L$ and $\am(\tilde{x}^\pm|\kappa^2)$ is the Jacobi amplitude function \cite{Abramowitz-Stegun1972}. Since $\d \am(\tilde{x}^\pm|\kappa^2)/\d \tilde{x}^\pm = \dn(\tilde{x}^\pm|\kappa^2)$, where dn is another Jacobi elliptic function, the Jacobian of the transformation \eqref{inverse-Lorentzian} is
\[\abs{\frac{\pd(x_0^+,x_0^-)}{\pd(x^+,x^-)}} = \frac{4 K^2\lp \kappa^2\rp}{\pi^2} \dn\lp \tilde{x}^+\middle|\kappa^2\rp \dn\lp \tilde{x}^-\middle|\kappa^2\rp.\]
The general tuned rectangle one-point function is thus
\bal
\ev{\Phi(t,x)}_{L/h} &= \abs{\frac{\pd(x_0^+,x_0^-)}{\pd(x^+,x^-)}}^{\Delta/2} \ev{\Phi(t,x)}_0 \non\\
&= C_\Phi\lb \frac{2K\lp \kappa^2\rp}{L} \frac{\sqrt{\dn \tx^+\dn \tx^-}}{\sn\tx^+\cn\tx^- +\sn \tx^-\cn\tx^+}\rb^\Delta,
\label{one-point}
\eal
where all of the Jacobi elliptic functions are abbreviated $\pq z = \pq\lp z\middle| \kappa^2\rp$ and where we used the identities $\sin[\am(z|k)] = \sn(z|k)$ and $\cos[\am(z|k)] = \cn(z|k)$. The result \eqref{one-point} is plotted in figure \ref{fig:one-point} for several values of $L/h$. One can show that
\eqref{one-point} is periodic with period $L$ by using the identities \cite{Whittaker-Watson1927}
\bal
\sn\lb z \pm K\lp \kappa^2\rp \middle| \kappa^2\rb &= \pm \frac{\cn\lp z \middle| \kappa^2\rp}{\dn\lp z \middle| \kappa^2\rp}\non\\
\cn\lb z \pm K\lp \kappa^2\rp \middle| \kappa^2\rb &= \mp\sqrt{1-\kappa^2}\frac{\sn\lp z \middle| \kappa^2\rp}{\dn\lp z \middle| \kappa^2\rp}\non\\
\dn\lb z \pm K\lp \kappa^2\rp \middle| \kappa^2\rb &= \frac{\sqrt{1-\kappa^2}}{\dn\lp z \middle| \kappa^2\rp}\non.
\eal

\begin{figure}
\bc
\includegraphics[scale=.4]{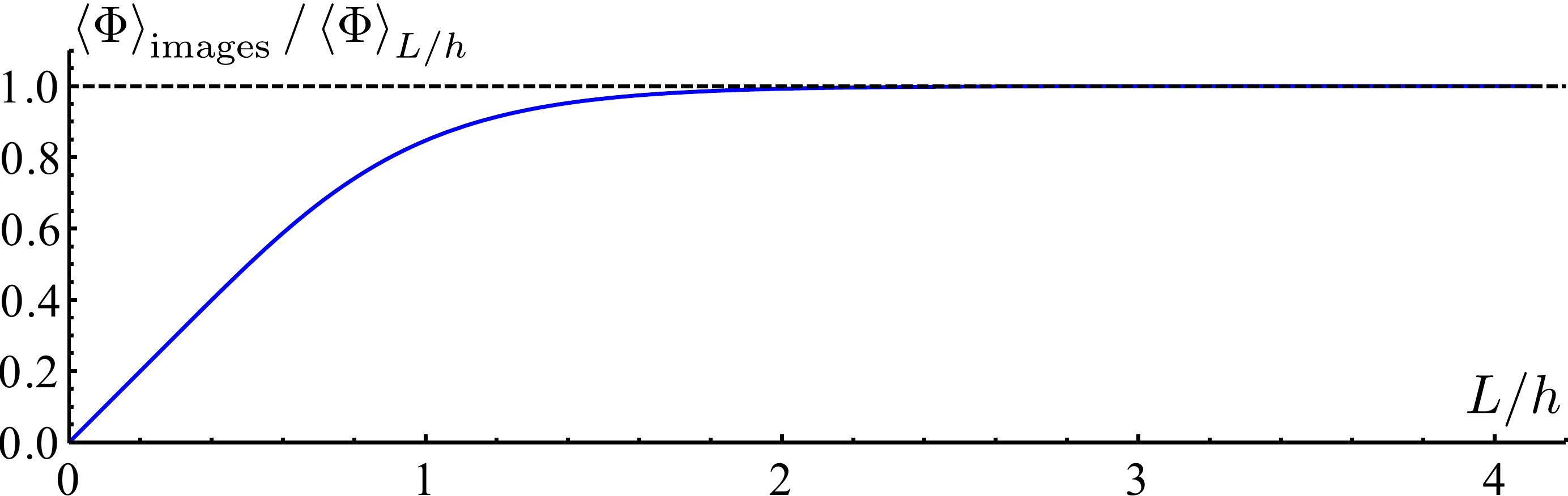}
\caption{Ratio of \eqref{cardy-one-point} to \eqref{one-point} evaluated at $x=L/2$ with $C_\Phi=1$. \eqref{cardy-one-point} is truncated with $N=20$ as is described in the text. Note that this ratio is time independent.}
\label{fig:one-point-compare}
\ec
\end{figure}

For comparison, the one-point function $\ev{\Phi(t,x)}_{L/h}$ was computed in \cite{Cardy2014} for $x= L/2$ and $L/h \gg 1$ by noting that it is a universal function of $L$, $h$, and $\Delta$ and so can be found by taking $\Phi$ to be the exponential of a massless scalar field. Using the method of images and the condition $L/h \gg 1$ to replace the image sum by an integral yields
\beq
\ev{\Phi(t,L/2)}_{\t{images}} \approx \lp \frac{\pi}{h}\rp^\Delta \prod_{m=-\infty}^\infty \lc \frac{\cosh \lb \frac{\pi L}{h}(t/L-m-\frac{1}{2})\rb \cosh \lp \frac{\pi L}{h}m\rp}{\cosh \lb \frac{\pi L}{h}(t/L-m)\rb \cosh \lb \frac{\pi L}{h}(m+\frac{1}{2})\rb}\rc^\Delta.
\label{cardy-one-point}
\eeq
The ratio of \eqref{cardy-one-point} to \eqref{one-point} --- evaluated at $x=L/2$ with $C_\Phi=1$ --- is plotted in figure \ref{fig:one-point-compare}.\footnote{The series given by the log of \eqref{cardy-one-point} fails to converge absolutely.  As a result, one must use care in taking the limit of an infinite number of terms. Figure~\ref{fig:one-point-compare} approximates \eqref{cardy-one-point} using
\[\ev{\Phi(t,L/2)}_{\t{images}} = \lp \frac{\pi}{h}\rp^\Delta \prod_{m=-N}^{N-1} \lc \frac{\cosh \lb \frac{\pi L}{h}(t/L-m-\frac{1}{2})\rb}{\cosh \lb \frac{\pi L}{h}(t/L-m)\rb}\rc^\Delta \prod_{m=-N}^{N}\lc \frac{\cosh \lb \frac{\pi L}{h}(m+\frac{1}{2})\rb}{\cosh \lp \frac{\pi L}{h}m\rp}\rc^\Delta\] for $N=20$.  This approximation explicitly preserves the desired symmetry $t \rightarrow -t$.  More naive truncations that do not explicitly respect this symmetry still converge as $N \rightarrow \infty$, but the limit is neither periodic nor consistent with time-reversal symmetry.}
As desired, the two agree for $L/h\gg 1$, though \eqref{cardy-one-point} clearly vanishes as $h \rightarrow \infty$ due to the explicit factor $(\pi/h)^\Delta$.  This factor was not stated explicitly in \cite{Cardy2014}, though we obtained it from \cite{private} and verified the result.

We may also check our result in the limit $L\to\infty$ ($\kappa \to 1$). From \eqref{rectangle-dimensions} we find $K(1)/L=2K(0)/h=\pi/h$. Using $\sn(z|1) = \tanh z$ and $\dn(z|1) = \cn(z|1)=\sech z$ we thus find
\[\lim_{L\to\infty}\ev{\Phi(t,x)}_{L/h} = C_\Phi\lb \frac{\pi}{h} \sqrt{\frac{\cosh 2\pi t/h + \cosh 2\pi x/h}{2}}\sech \frac{\pi t}{h} \csch\frac{\pi x}{h}\rb^\Delta.\]
Far from the $x=0$ boundary this becomes
\beq
\lim_{L\to\infty}\ev{\Phi(t,x \gg h)}_{L/h} = C_\Phi\lp \frac{\pi}{h} \sech \frac{\pi t}{h}\rp^\Delta,
\label{thermal-one-point}
\eeq
which yields $C_\Phi(2\pi/h)^\Delta \e^{-\Delta\pi t/h}$ for  $\pi t\gg h$. This coincides with the decay found for infinite intervals in \cite{Calabrese2006, Calabrese2007}.

We may also compute the expectation values of the stress tensor and the energy in tuned rectangle states. As mentioned above, we will begin on the half-plane and obtain the stress tensor on the tuned Euclidean rectangle by the Euclidean conformal transformation \eqref{map}.  We will focus on the stress tensor at $t=0$ for which Wick-rotation to Lorentz signature is essentially trivial: $\ev{T_{tt}}_{L/h} =-\ev{T_{\tau\tau}}_{L/h}$, and time-reversal symmetry requires $\ev{T_{tx}}_{L/h} =0$.

On the half-plane the symmetries require $\ev{T_{\mu \nu}}_{\t{HP}}=0$.  In particular, translation and reflection symmetry in $y_2$ implies $\ev{T^{y_1y_2}}_{\t{HP}} = \ev{T^{y_2y_1}}_{\t{HP}}=0$.\footnote{This condition ensures that there is no energy flux across the boundary.} Since the stress tensor has conformal weight 2, conformal invariance of $\ket{B}$ and the vanishing trace on the plane require $\ev{T^{y_1y_1}}_{\t{HP}} = -\ev{T^{y_2y_2}}_{\t{HP}} = C/y^2$ for some constant $C$. But the conservation law $\pd_\nu \ev{T^{\mu\nu}}_{\t{HP}} = 0$ forces $C=0$.

The stress tensor on the rectangle thus comes only from the anomalous term in the conformal transformation law for $\ev{T_{\mu \nu}}$.  Since the trace vanishes on the rectangle as well, we have
\beq
\ev{T_{\tau\tau}}_{L/h} = -\ev{T_{ww}}_{L/h} - \ev{T_{\bar{w}\bar{w}}}_{L/h} = \frac{1}{\pi}\ev{\re T(w)}_{L/h},
\label{stress-conventions}
\eeq
where $T(w) = -2\pi T_{ww}$. Under a general conformal transformation $z\to w(z)$, the stress tensor transforms as \cite{DiFrancesco-Mathieu-Senechal1997}
\beq
\ev{T(w)}_{L/h} = \lb z'(w)\rb^2 \ev{T(z)}_{\t{HP}} + \frac{c}{12}\{z;w\},
\eeq
where
\[\{z;w\} \equiv \frac{z'''(w)}{z'(w)} - \frac{3}{2}\lb \frac{z''(w)}{z'(w)}\rb^2\]
is the Schwarzian derivative.

\begin{figure}
\bc
\includegraphics[scale=.4]{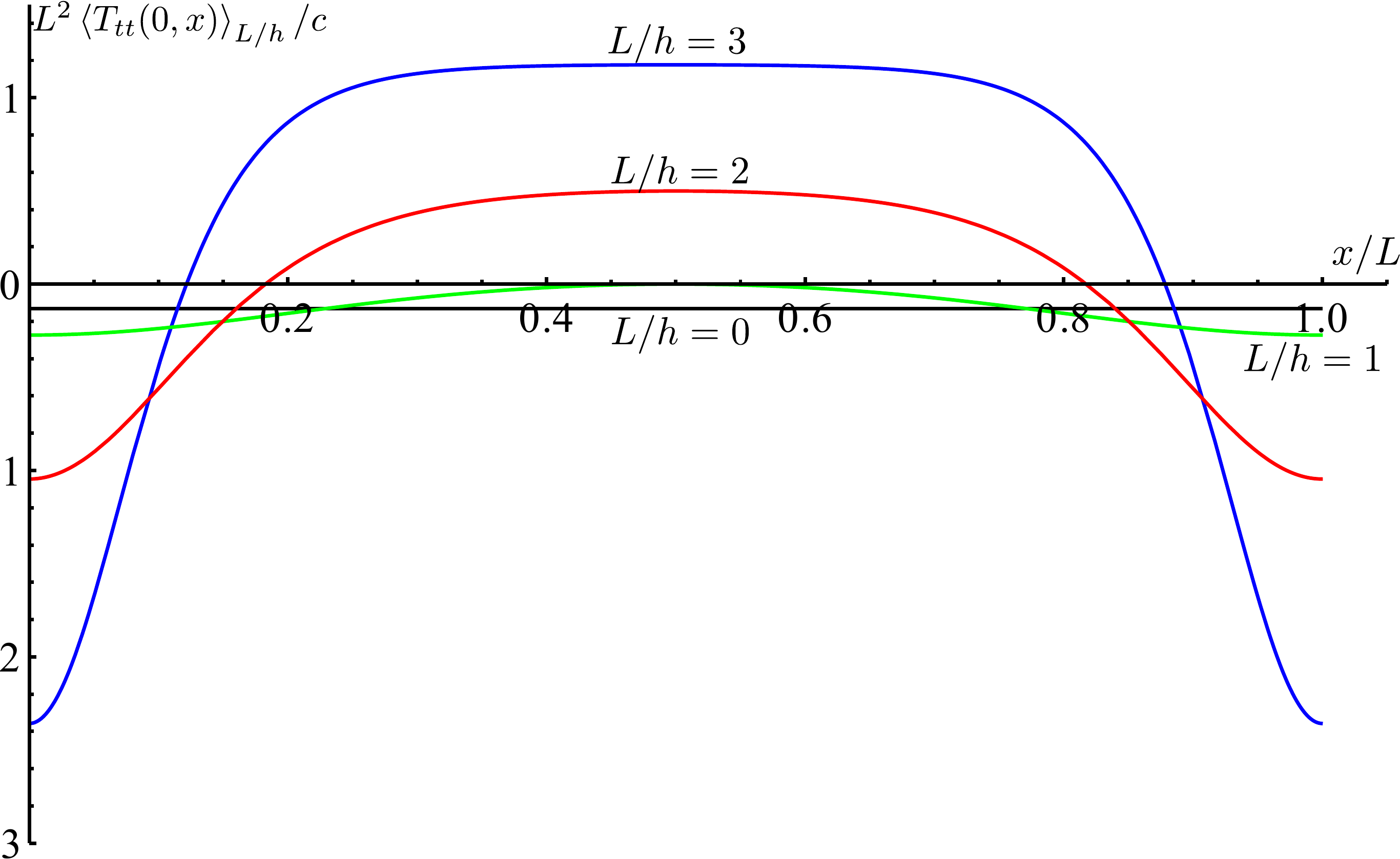}
\caption{Energy density as a function of distance along the Lorentzian strip at $t=0$.}
\label{fig:energy-density}
\ec
\end{figure}

The stress tensor for $h = \infty$ follows from \eqref{grnd-map}. One finds $\{z;w\} = \pi^2/2L^2$, so the energy of the ground state is just
\beq
\ev{E}_0 = -\frac{c\pi}{24L}.
\label{grnd-energy}
\eeq
The stress tensor for an arbitrary tuned rectangle follows from  \eqref{map}:
\beq
\ev{T(w)}_{L/h} = \frac{c}{24L^2} K^2\lp \kappa^2\rp \frac{3\dn^4\lp \tilde{w}\middle|\kappa^2\rp - \lp 2-\kappa^2\rp \dn^2\lp \tilde{w}\middle|\kappa^2\rp + 3\lp 1-\kappa^2\rp}{\dn^2\lp \tilde{w}\middle|\kappa^2\rp},
\label{stress}
\eeq
where $\tilde{w}\equiv w K\lp \kappa^2\rp/L$. We plot the energy density at $t=0$ in figure \ref{fig:energy-density} for several values of $L/h$. States with non-zero values of $L/h$ ($\kappa<1$) have energy densities that peak near the center of the strip and have negative dips near the ends of the strip. As $L/h\to \infty$ ($\kappa \to 1$), the central peak becomes a wide plateau with a value at the center of the strip that diverges as $h^{-2}$ and becomes independent of $L$:
\beq
\ev{T_{tt}(0, L/2)}_{L/h\to\infty} = \frac{c\pi}{24h^2} + \mathcal{O}\lp\sqrt{1-\kappa^2}\rp.
\label{stress-short}
\eeq
Note that this is the $tt$ component of the thermal stress tensor at inverse temperature $\beta = 2h$.

Finally, consider the Wick rotation $w\to x - t$, $\bar w\to x + t$.  Since a shift in $t$ by an integer multiple of $L$  ($t\to nL$) shifts the arguments of the Jacobi elliptic functions by $nK\lp \kappa^2\rp$, the Lorentz-signature stress tensor is periodic with period $L$ as expected.

\section{Conformal Transformations to the Thermal Minkowski Plane}
\label{sec:minkowski}

We noted above that tuned rectangle states with arbitrary $L$, $h$ are Lorentz-signature conformal transformations of the ground state on the strip of width $L$.  As a result, these states are periodic in time with period $L$ and do not thermalize.  Recall, however, that \cite{Calabrese2006, Calabrese2007, Calabrese2007b,Hartman2013b} used conformal boundary states on the infinite Minkowski plane to model the thermalization of generic states.  Such Minkowski states are the $L \rightarrow \infty$ limit of tuned rectangles with fixed $h$.  This suggests that there may be a sense in which Minkowski conformal boundary states fail to thermalize as well.

We make this precise below by showing that Minkowski conformal boundary states are again conformally related to the ground state on the finite-$L$ strip with the associated boundary condition $B$.  Roughly speaking, the desired conformal transformation is the composition of the singular $h \rightarrow 0$ ($\kappa\to1$) limit of \eqref{Lorentzian-map} with an infinite scale transformation that takes $L \rightarrow \infty$, but it is easier to derive the correct map from Euclidean considerations.  We will find that our map takes the Minkowski plane to the diamond shown in figure \ref{fig:map}, which we call the fundamental diamond.  In particular, we see that the states $|h \rangle_{\t{Mink}} \propto \e^{-\frac{h}{2}H}|B\rangle$ defined on the Minkowski plane admit extensions of their time evolution past future null infinity ($\mathscr{I}^\pm$).  Since this extension is given by the ground state on the strip, it is in no way thermal.  The lack of periodicity and apparent thermalization in \cite{Calabrese2006, Calabrese2007, Calabrese2007b,Hartman2013b} can be viewed as being due to confining attention to the fundamental diamond, whose extent in time is too small to feel the influence of the boundaries of the strip in figure \ref{fig:map}.

We also show below that more general states on the Minkowski plane cannot be smoothly extended in the same way.  This raises the question of whether these more general Minkowski states might ``thermalize" in a deeper sense than do conformal boundary states.

We now construct the desired conformal map. As in section \ref{sec:cft-states}, we begin with the Euclidean conformal transformation and then Wick rotate the result.  A useful first step in constructing the desired transformation is to relate the path integral over the (say, eastern) half of a unit sphere to those defining both the Minkowski states $|h\rangle_{\t{Mink}}$ and the ground state $|0\rangle_{\t{strip}}$ on the strip of length $L$.  Taking $\theta$ to be the usual polar angle based at the north pole, our hemisphere is $\theta, \phi \in [0,\pi]$.  The conformal map to the (Euclidean) $h = \infty$ strip with finite $L$ and Cartesian coordinates $X, T_E$ is defined by $T_E = \frac{L}{\pi} \ln \tan \theta/2 , X = \frac{L\phi}{\pi}$.  The map to the (Euclidean) $L = \infty$ strip with finite $h$ and Cartesian coordinates $x, \tau$ is similar, differing only by a $\pi/2$ rotation of Euclidean space and a shift of the origin.  The full form of this map is rather complicated in terms of $\theta, \phi$, though it simplifies at $\tau =0$ ($\theta = \pi/2)$ where we find $x = \frac{h}{\pi} \ln \tan \phi/2$. So for $\tau = T_E = 0$ we have  $x = \frac{h}{\pi} \ln \tan (\pi X/2L)$.  Time-reversal symmetry then implies that the Wick-rotated conformal transformation can be written in terms of null coordinates $U=T-X$, $V=T+X$ on the fundamental diamond  and $u= t-x$, $v=t+x$ on the Minkowski plane as
\beq
\label{UV}
U = -\frac{2L}{\pi} \arctan \e^{-2\pi u/\beta} \qquad \t{ and } \qquad V = \frac{2L}{\pi} \arctan \e^{2\pi v/\beta},
\eeq
where we have introduced the parameter $\beta = 2h$, which, from \eqref{stress-short}, has the interpretation of inverse temperature for $L\to\infty$. As advertised, this maps the Minkowski plane to the fundamental diamond $-U,V\in[0,L]$. The map \eqref{UV} relates the metrics on the Minkowski plane and the fundamental diamond by $\d s^2_{\t{Mink}} = \Omega^2\; \d s^2_{\t{diam}}$ with the conformal factor
\beq
\Omega = \frac{\beta}{2L} \sqrt{\cosh\frac{2\pi u}{\beta} \cosh \frac{2\pi v}{\beta}}.
\eeq

The map \eqref{UV} is familiar from the study of systems \textit{without} boundaries, where it is well-known for mapping  a diamond of width $L$ in a cylinder of circumference $2L$ to the thermal Minkowski plane (see \cite{Hislop:1981uh,Haag:1992hx}, or more explicitly \cite{Maldacena:1998bw}). In other words, applying this map to the stress tensor in the thermal state on Minkowski space (with $\ev{T_{tt}(t,x)}_{\t{Mink}} = \ev{T_{xx}(t,x)}_{\t{Mink}} = \pi c/6\beta^2$) yields the stress tensor $\ev{T_{tt}(t,x)}_{\t{diam}} = \ev{T_{xx}(t,x)}_{\t{diam}}= -\pi c/24L^2$ on the fundamental diamond, which is characteristic of the ground state on a cylinder of circumference $2L$.  Since $\ev{T_{\mu \nu}}_{\t{HP}} =0$ on the half-plane with conformal boundary conditions, expectation values of the stress tensor in $|0\rangle_{\t{strip}}$ with width $L$ coincide with those in the ground state of the cylinder of circumference $2L$.  For the same reason, the expectation value of the stress tensor in $|h\rangle_{\t{Mink}}$ is precisely thermal.  Thus \eqref{UV} relates these stress tensors as desired. It is also straightforward to check \eqref{UV} by examining the one-point functions of primary operators.  Applying \eqref{UV} to the one-point function $\ev{\Phi}_{\t{Mink}}$ in $|h\rangle_{\t{Mink}}$ given by \eqref{thermal-one-point} yields
\beq
\ev{\Phi}_{\t{diam}} = \Omega^\Delta \ev{\Phi}_{\t{Mink}} = C_\Phi \lb \frac{\pi}{L}\sqrt{\cosh \frac{2\pi u}{\beta} \cosh \frac{2\pi v}{\beta}} \sech \frac{\pi(u+v)}{\beta}\rb^\Delta
\label{1pt-diam},
\eeq
which agrees with \eqref{1p-rect} in the fundamental diamond.

The strip ground state $|0\rangle_{\t{strip}}$ thus provides a smooth conformal extension of $|h\rangle_{\t{Mink}}$ past $\mathscr{I}^\pm$.  This is a special property of conformal boundary states not shared by more general states on the Minkowski plane. Indeed, we now describe a class of states which admit no smooth conformal extensions past $\mathscr{I}^\pm$.

For simplicity, we assume that the theory admits a $U(1)$ symmetry with generator $Q$ such that $Q|h \rangle_{\t{Mink}} =0$. The desired class of states may be obtained from $|h \rangle_{\t{Mink}}$ by acting at $t=0$ with a unitary operator ${\cal O}$ that is invariant under the combined action of translations $x \rightarrow x + \lambda$ and $\e^{\i p\lambda Q}$ for some momentum $p$ and any displacement $\lambda$.  If ${\cal O}$ has non-trivial commutator with some complex primary $\Phi$ with charge $q\neq 0$ under the above $U(1)$, the expectation value of $\Phi$ in ${\cal O} |h\rangle_{\t{Mink}}$ will be of the form $\langle \Phi (t=0) \rangle = D_\Phi \e^{-\i qpx}$.  We assume $D_\Phi \neq 0$.  On the other hand, the expectation value of the stress tensor remains translation invariant.  If ${\cal O}$ also preserves the $\mathbb{Z}_2$ symmetry that acts simultaneously as parity ($x \rightarrow -x$) and charge conjugation ($Q \rightarrow -Q$), then by making an additional scale transformation we can take the stress tensor to have the same (thermal) expectation value as in $|h\rangle_{\t{Mink}}$.  This forces any conformal map that might define a conformal extension of our state to have the form \eqref{UV} near $\mathscr{I}^\pm$.

However, there are cases in which one can show that the expectation value of $\Phi$ at late times is of the form
\beq
\ev{\Phi}_{\t{Mink}} = \frac{\tilde{D}_\Phi}{2} \lb \frac{2\pi}{\beta} \sech \frac{\pi(u+v)}{\beta}\rb^\Delta \lp \e^{\i pqu} + \e^{-\i pqv}\rp
\label{1pt-diamp}.
\eeq
This result may be motivated by supposing that the real and imaginary parts of $\Phi$ do not interact significantly at late times and using a conformal transformation to generate e.g. $\re \ev{\Phi}_{\t{Mink}}$ from \eqref{thermal-one-point}.  In the holographic context it follows from the known spectrum of quasi-normal modes on the BTZ black hole as described in appendix \ref{sec:holographic}.

On the fundamental diamond $\ev{\Phi}_{\t{diam}} = \Omega^\Delta \ev{\Phi}_{\t{Mink}}$ is thus of the form \eqref{1pt-diam} multiplied by $\e^{\i pqu} + \e^{-\i pqv}$. It is this final factor that causes the trouble.  The future edges of the diamond are defined by either $u \rightarrow +\infty$ or $v \rightarrow +\infty$.  There the limits of \eqref{1pt-diam} are smooth and non-zero.  But since the factor $\e^{\i pqu} + \e^{-\i pqv}$ oscillates infinitely many times as these edges are approached, our new state fails to be smooth there and admits no smooth conformal extensions beyond the diamond.

One might nevertheless wonder if the CFT dynamics allows one to evolve the state through the resulting singularity.  In other words, might our new state admit a unique and well-defined but singular conformal extension to the timelike strip?  The smooth expectation value of the stress tensor may be taken as an indication that it does, though a complete answer lies beyond the scope of the present work.

\section{Discussion}
\label{sec:discussion}

We have shown that the tuned rectangle states of \cite{Cardy2014} are Lorentz-signature conformal transformations of the ground state on the strip. This immediately explains the periodicity and lack of thermalization observed in \cite{Cardy2014}. It seems clear that the exact periodicity of the analogous local quench studied in \cite{local} is similarly due to the final state being a Lorentz-signature conformal transformation of the ground state.

We also showed that the related states $|h\rangle_{\t{Mink}}$ used to model thermalization in \cite{Calabrese2005, Calabrese2006, Calabrese2007, Calabrese2007b,Calabrese:2009ez, Hartman2013b} can be conformally mapped to the ground state of the finite $L$ strip restricted to the fundamental diamond.  The strip ground state thus provides a smooth conformal extension of such states beyond future infinity of the Minkowski plane.  This extension is distinctly non-thermal.  In contrast, we argued that more general states admit no smooth conformal extension.

The actual conformal transformation used to relate $|h\rangle_{\t{Mink}}$ to the fundamental diamond in $|0\rangle_{\t{strip}}$ is precisely that used to relate a thermal state on the Minkowski plane to a diamond on the (half) cylinder. It is thus related at the formal level to the fact that the cylinder vacuum provides a conformal extension of the Minkowski thermal state. Recall, however, that the thermal state is highly mixed, and the cylinder vacuum extension purifies the state by adding additional degrees of freedom.  In contrast, the states $|h\rangle_{\t{Mink}}$ are already pure and --- because the strip boundaries pass through the right and left corners of the fundamental diamond (see figure \ref{fig:map}) --- the extension to $|0\rangle_{\t{strip}}$ adds no new degrees of freedom.  The only additional information supplied beyond that already in $|h\rangle_{\t{Mink}}$ is the particular choice of boundary condition imposed at the edges of the strip.

This observation provides a sense in which $|h\rangle_{\t{Mink}}$ retains certain non-thermal characteristics even in the far future. We noted that more general states admit no smooth conformal extensions, though we left open the question of whether they might admit well-defined but non-smooth conformal extensions with similar non-thermal characteristics.

It is interesting to ponder the implications for general studies of thermalization. As an example, \cite{Calabrese2005, Calabrese2006, Calabrese:2009ez} argued that the mutual information of two equal-sized length $\ell$ intervals in $|h\rangle_{\t{Mink}}$ first grows to reach a thermal value and then displays a remarkable dip at the time $t =\ell + D$ where $D$ is the distance between the intervals.  Here as usual we set the speed of light to one, so this is the time when quasi-particles moving to the right at the speed of light that originally filled the left interval would be entirely contained in the right interval. This was a key ingredient in the arguments of \cite{Calabrese2005, Calabrese2006, Calabrese:2009ez} for a simple quasi-particle description of entanglement (though see \cite{Cardy2014} for a more refined picture). While we understand that \cite{HTA} will question whether this result can hold outside the rather special framework of rational coformal field theories,\footnote{Other aspects of the quasi-particle picture were questioned in \cite{Andrade:2013rra}, noting certain tensions with holographic calculations. See also related comments in \cite{Asplund:2013zba}.} if confirmed for general CFTs the above dip would be an example of the kind of late-time ($t \gg h$) non-thermal behavior we have in mind.  We would then be led to ask whether such results might be intimately tied to the existence of conformal extensions, which might be particular to the states $|h\rangle_{\t{strip}}$, or at least to states in 1+1 CFTs.

\acknowledgments

We thank John Cardy, Sebastian Fischetti, Thomas Hartman, William Kelly, Rob Myers, and Mark Srednicki for discussions. This work was supported in part by the National Science Foundation under Grant No.~PHY11-25915, by FQXi grant FRP3-1338, and by funds from the University of California.  In addition, K.K.\ is supported by the NSF GRFP under Grant No.~DGE-1144085.

\appendix

\section{Holographic Description}
\label{sec:holographic}

We now provide a few brief comments on the holographic description of conformal boundary states and show explicitly that the gravitational solutions dual to $\ket{L/h}_{\t{strip}}$ do not contain black holes for any value of $L/h$.  The main technical point is to justify the behavior \eqref{1pt-diamp} near $\mathscr{I}^\pm$.  However, we also find the holographic perspective useful for better understanding the comments of section \ref{sec:minkowski}.  We proceed quickly and refer readers unfamiliar with holography to e.g.\ \cite{Skenderis:2002wp,Fischetti2012} for reviews of the relevant technology. Previous holographic studies of quantum quenches include \cite{Hartman2013b, Takayanagi2010, Ugajin2013, Asplund2011, Asplund:2013zba, Basu2012, Basu2013, Buchel2013, Buchel2013b, Balasubramanian2011b, Balasubramanian2011c, Allais2012, Nozaki2013, Das2012, Liu2014, Abajo-Arrastia:2014fma}.

Before beginning, recall from \cite{Aharony:2010ay} that the holographic dual of a BCFT can often be described by adding appropriate orbifolds/orientifolds to the bulk that intersect the AdS conformal boundary at the boundary of the CFT.  In parallel with the Randall-Sundrum constructions \cite{Randall:1999ee,Randall:1999vf}, refs. \cite{Takayanagi2011,Fujita2011} introduced a simple phenomenological model in which one simply chooses a surface $\Sigma$ in the bulk (called the end-of-the-world brane below) on which one takes the bulk spacetime to end.  One then imposes boundary conditions requiring this surface to intersect the AdS conformal boundary at the boundary of the CFT.  One also requires the bulk fields to satisfy a boundary condition at the end-of-the-world brane that respects diffeomorphism invariance and which provides a good variational principle for the system.  We use this framework below and impose a Neumann boundary condition for the metric
\begin{equation}
\label{noK}
K_{\mu \nu} = 0,
\end{equation}
where $K_{\mu \nu}$ is the extrinsic curvature of $\Sigma$, and a Dirichlet boundary condition
\begin{equation}
\phi|_\Sigma = \phi_0,
\end{equation}
for some minimally-coupled bulk scalar $\phi$, which for simplicity we treat as free aside from its coupling to bulk Einstein-Hilbert $\t{AdS}_3$ gravity.  We take this scalar to have mass $m$ and the dual CFT operator $\Phi$ to have dimension $\Delta = 1 + \sqrt{1+m^2\ell^2}$, where $\ell$ is the AdS length. Here, for simplicity, we restrict to operators with $\Delta >1$. We also ignore any complications from internal compact dimensions or additional $\t{AdS}_3$ fields.

Before turning to the state $|h \rangle_{\t{Mink}}$, we first discuss the holographic description of the strip ground state $|0\rangle_{\t{strip}}$.
Recall that a Lorentz-signature strip of length $L$ and infinite height is conformally equivalent to global $\t{AdS}_2$, whose line element may be written
\begin{equation}
\label{AdS2}
\d s^2_{\t{AdS}_2} = \ell^2 \sec^2 \theta \lp -\d t^2 + \d\theta^2 \rp
\end{equation}
with $\t{AdS}_2$ length scale $\ell$ the same as the $\t{AdS}_3$ length scale. Here we take $t\in (-\infty,\infty)$ and $\theta\in [-\pi/2,\pi/2]$. Since we expect the strip ground state to be invariant under the $SO(1,2)$ isometries of \eqref{AdS2}, the holographic dual spacetimes should admit foliations by $\t{AdS}_2$ slices with $\phi$ constant on each slice.

Choosing a conformal frame where the boundary metric is \eqref{AdS2}, the Fefferman-Graham coordinate $z$ will be constant on each slice. The bulk metric is then (using $\Delta >1$)
\begin{align}
\label{fefferman-graham}
\d s^2 = \frac{\ell^2}{z^2}\lb \d z^2 + \gamma_{\mu\nu}(x,z) \d x^\mu \d x^\nu\rb ; \cr
\gamma_{\mu\nu}(x,z) = \gamma^{(0)}_{\mu\nu}(x) + z^2 \gamma^{(2)}_{\mu\nu}(x) + \dots,
\end{align}
where the radial Hamiltonian constraint requires
\beq
\gamma^{(0)\mu\nu}\gamma^{(2)}_{\mu\nu} = -\frac{1}{2}R.
\label{constraint}
\eeq
Here $z$ is a radial coordinate that approaches $0$ near the AdS boundary and $R$ is the Ricci curvature scalar of the boundary metric $\gamma^{(0)}_{\mu\nu}$ which we take to be \eqref{AdS2}.   Since AdS${}_2$ symmetry requires $\gamma^{(2)}_{\mu\nu} \propto \gamma^{(0)}_{\mu\nu}$, we readily solve \eqref{constraint} to find $\gamma^{(2)}_{\mu\nu} = 2 \gamma^{(0)}_{\mu\nu}$.

Thus, to this order the solution is independent of $\phi_0$.  Since the boundary stress tensor \cite{Henningson:1998gx,Balasubramanian:1999re} is determined by $\gamma^{(0)}_{\mu\nu}$ and $\gamma^{(2)}_{\mu\nu}$ we see that it is also independent of $\phi_0$.  A detailed calculation confirms that it agrees with the expectation value in $|0\rangle_{\t{strip}}$ found in section \ref{sec:one-point}.

It is now straightforward to construct holographic duals to more general tuned rectangle states using the observation of section \ref{sec:construction} that they are Lorentz-signature conformal transformations of $|0\rangle_{\t{strip}}$.  CFT conformal transformations are implemented in the bulk by appropriate diffeomorphisms, so solutions dual to any $|L/h\rangle_{\t{strip}}$ describe the same geometry as for $|0\rangle_{\t{strip}}$.  All that changes is the choice of conformal compactification (i.e., the choice of a particular Fefferman-Graham radial coordinate $z$).  This immediately explains the observation of footnote \ref{noBH} that, despite having high energy for large $L/h$, such solutions do not form black holes.

\begin{figure}
\bc
\includegraphics[scale=1]{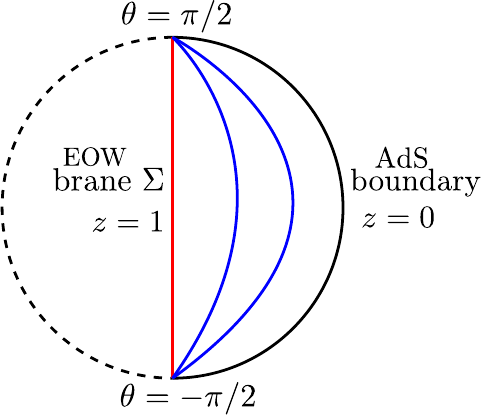}
\caption{A constant time slice of the bulk ground state. The red line denotes the end-of-the-world brane $\Sigma$ at $z=1$, and the two blue lines indicate $\t{AdS}_2$ slices of constant $z$.  In the limit $\phi_0 \rightarrow 0$ the metric becomes that of empty global $\t{AdS}_3$.}
\label{fig:bulk-grnd}
\ec
\end{figure}

One may get some feel for the solutions by taking $\phi_0$ small enough that gravitational back-reaction can be ignored.  The solution must then be just some region of AdS${}_3$ whose boundary respects the symmetries and the condition \eqref{noK}.  The obvious choice is to use the AdS${}_2$ slicing of AdS${}_3$, with $\Sigma$ the AdS${}_2$ of minimal radius.  Our solution is then just half of global AdS${}_3$ as shown in figure \ref{fig:bulk-grnd}.   One may similarly construct interesting excited states dual to CFTs on a strip by using half of any (global) BTZ black hole \cite{Banados1993}, though these break the $SO(1,2)$ symmetries down to just time translations \cite{Takayanagi2011}.

We now return to general $\phi_0$. We wish to discuss the limit that the length $L$ of the rectangle becomes infinite.  As described in section \ref{sec:minkowski}, the associated states $|h \rangle_{\t{Mink}}$ are conformally related to the fundamental diamond of $|0\rangle_{\t{strip}}$.  This diamond defines an associated region of the conformal boundary of any bulk solution.  The intersection in the bulk of the causal past and future of this diamond defines a region analogous to the Rindler wedge of global AdS${}_3$.

Indeed, there is a natural set of Rindler-like coordinates defined on this wedge as follows.  We use a Fefferman-Graham gauge in which the boundary metric in the fundamental diamond becomes the usual 1+1 Minkowski metric.  Noting that spatial translations of this metric coincide with one of the original AdS${}_2$ isometries, we see that the associated radial coordinate $z_R$ defines a foliation in which every slice continues to be invariant under translations.  We may thus use the Killing coordinate $x_R$ associated with this isometry and a time coordinate $t_R$ that is orthogonal to $x_R$ in each slice.  The remaining scaling freedom is fixed by associating a temperature $1/\beta$ with our fundamental diamond as in section \ref{sec:minkowski} and by taking $x_R$ and $t_R$ to be normalized with respect to the boundary metric defined by $z_R$.  Note that our boundary conditions require the solution in this Rindler-like region to asymptote to the usual Rindler wedge of empty global AdS${}_3$ as $t_R \rightarrow \pm \infty$.

Of course, another name for the Rindler wedge of global AdS${}_3$ is the (exterior of the) planar BTZ black hole.\footnote{This statement is well-known, though it is difficult to find a reference which describes the details in this language. By {\it planar} BTZ black hole, we mean the universal cover of the black hole \cite{Banados1993} given by removing the identifications of the spatial coordinate.  The relevant coordinate transformations relating this to the standard global AdS${}_3$ metric are described in \cite{Banados1993}. The planar BTZ black hole contrasts with the global BTZ black hole discussed above, in which the spatial coordinate is periodic.}  In the same way, when expressed in the above Rindler coordinates our solution for general $\phi_0$ defines the exterior of some dynamical BTZ-like black hole which we see must approach the usual BTZ solution in the far future and past.  Continuing our solution beyond the horizon\footnote{That $\Sigma$ lies either at or beyond the horizon follows \cite{Wall:2012uf,MVM} from the fact that the $t=0$ slice of $\Sigma$  is a minimal surface due to the boundary condition \eqref{noK}.  For more general boundary conditions this brane may lie outside the horizon at $t=0$, though the boundary conditions at $z=0$ require it to fall behind the horizon at some $t_R$.} one will of course find that the spacetime ends at the end-of-the-world brane $\Sigma$.  For $\phi_0 = 0$ we obtain the solution used to describe holographic duals of the states $\ket{h}_{\t{Mink}}$ in \cite{Hartman2013b}, which studied the resulting entanglement properties in detail.

This is the holographic description of the conformal map from section \ref{sec:minkowski} relating $|0\rangle_{\t{strip}}$ and $|h\rangle_{\t{Mink}}$.  The bulk version of the conformal boundary state $|h\rangle_{\t{Mink}}$ is just a particular region of the spacetime dual to $|0\rangle_{\t{strip}}$.  The thermal properties are due to confining attention to the fundamental diamond and ignoring the region farther to the future.  Though this region may be legitimately called a (planar) black hole, the solution may be extended to the future as a spacetime with no black hole.  Furthermore, although the end-of-the-world brane is in a sense hidden behind the black hole horizon for purposes associated with observations in the fundamental diamond, it is directly visible to observers whose worldlines extend farther in time.

The late-time behavior \eqref{1pt-diamp} is now justified for holographic theories by noting that our solution becomes the usual BTZ black hole as $t_R \rightarrow \infty$ and recalling \cite{Birmingham2001} that, with momentum $p$,  the lowest scalar quasi-normal modes of BTZ have frequencies
$\omega = \pm p - 2\pi\i\Delta/\beta$.

\bibliography{thermalization}
\bibliographystyle{JHEP}

\end{document}